\documentclass[
    amsmath,
    amssymb,
    reprint,
    aps,
    prx,
    superscriptaddress,
    longbibliography
]{revtex4-1}

\usepackage[utf8]{inputenc}
\usepackage[T1]{fontenc}
\usepackage{physics}
\usepackage{nicefrac}
\usepackage{bbm, bm}
\usepackage{makecell}
\usepackage{multirow}
\usepackage{booktabs}
\usepackage{overpic}
\usepackage{tikz}
\usepackage{pgfplots}
\pgfplotsset{compat=1.18}
\usepackage[dvipsnames]{xcolor}

\usepackage[breaklinks=true, colorlinks]{hyperref}
\hypersetup{linkcolor=blue, citecolor=blue, urlcolor=MidnightBlue}
\usepackage{cleveref}

\begin{document}

\title{Truncated Wigner dynamics of biclique quantum spin glasses}

\author{Dries Sels}
\affiliation{Department of Physics, Boston University, 590 Commonwealth Ave., Boston, Massachusetts 02215, USA}
\affiliation{Center for Computational Quantum Physics, Flatiron Institute, 162 5th Avenue, New York, NY 10010, USA}

\date{\today}

\begin{abstract}
Quantum spin glasses are often considered testbeds for studying quantum optimization algorithms and as such have been the subject of various quantum advantage claims. Here we investigate the near adiabatic dynamics of biclique quantum spin glasses within the (discrete) truncated Wigner approximation (TWA). Benchmarks on small systems show that TWA recovers sample-to-sample fluctuations of the Edwards-Anderson order parameter, over a wide range of annealing times, with increasing fidelity when the system size increases. We extract critical exponents from the Binder cumulant in line with theoretical expectations, reproducing recent quantum experiments. The computational cost of the method is minimal and it can easily be applied to tens of thousands of qubits.  
\end{abstract}

\maketitle

\emph{Introduction --}~Many combinatorial optimization problems are naturally reformulated as ground state problems of Ising spin glasses, a perspective that has connected statistical physics to computer science and which has motivated both classical approaches and quantum algorithms. Quantum annealing promises to guide a spin glass toward equilibrium faster than thermal annealing, making it a prime target for quantum computing~\cite{albash18}. In this context D-wave Quantum inc. has demonstrated quantum critical spin-glass dynamics on hundreds to thousands of qubits, most notably in references~\cite{dwave23,dwave25}. In Ref.~\cite{dwave25} D-wave presents results on quantum quench dynamics on a variety of different Ising spin glasses, including spin models defined on 2D square lattice, 3D square and diamond lattices, and a biclique graph. After extensive comparison to state-of-the-art numerical methods these experiments were claimed to be beyond the reach of classical computation. Recent developments in classical computation, in particular the adoption of belief propagation techniques in tensor network methods, have however enabled more accurate and precise simulation of large quantum systems~\cite{alkabetz21,begusic24,Tindall24}. Recent results in Ref.~\cite{Tindall26} show that one can indeed efficiently capture the quantum critical dynamics of local Ising spin glasses, using belief propagation to keep up with the entanglement generated during the time evolution and then extracting expectation values with more sophisticated cluster expansions and boundary matrix product states. Some limitations of these methods have been brought up in subsequent comments~\cite{king2025commentondynamicsdisordered,commentscience26}.

The computational cost of tensor network methods increases with the connectivity of the graph, for tensors with bond dimensions $\chi$, living on a graph with connectivity $k$, typical operations scale like polynomials of $\chi^{k}$. The cost thus increases exponentially with the connectivity, making these methods (if applied directly) prohibitively expensive even when the bond dimension is small. Biclique graphs have extensive connectivity and are thus formally out of reach of the techniques presented in Ref.~\cite{Tindall26}. It has recently been claimed that, \emph{on that point alone, D-wave's supremacy claim stands unchallenged}~\cite{KingLinked26,dwavequantumDWavex2019sQuantum}. The goal of this short paper is to address this point in particular. Is quantum annealing of extensively connected Ising spin glasses a viable route to quantum advantage? In particular, can one efficiently simulate the biclique spin glass problem presented in Ref.~\cite{dwave25}? 
\begin{figure}[t]
    \centering
    \includegraphics[width=0.8\linewidth]{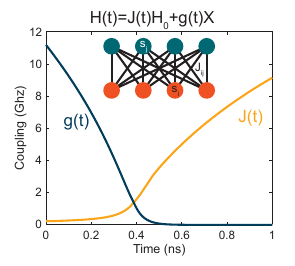}
    \caption{Quantum annealing schedule, with transverse field $g(t/t_a)$ and overall Ising coupling $J(t/t_a)$, for an annealing time of $t_a=1$ns. The schedule is used in D-wave's quantum annealing experiment~\cite{dwave25}. The inset shows a $K_{4,4}$ biclique graph.}
    \label{fig:diagram}
\end{figure}
In what follows I will provide evidence for the proposition that, in the large system size limit, the universal dynamics of biclique quantum spin glasses -- including sample-to-sample fluctuations -- can be capture by the truncated Wigner approximation (TWA). I will define the problem, set up the method, and after benchmarking the method, I will end with a brief discussion of the implications. 

\emph{Problem --}Consider time-dependent Ising Hamiltonians of the form 
\begin{eqnarray}
    &&H(t)=J(t/t_a) H_0 + g(t/t_a) H_x,  \nonumber \\
    &&H_0= \sum_{ij}J_{ij} \sigma^z_i \sigma^z_j, \, {\rm and}\ \, H_x=-\sum_i \sigma^x_i,
\end{eqnarray}
\begin{figure*}[t]
    \centering
    \includegraphics[width=0.98\linewidth]{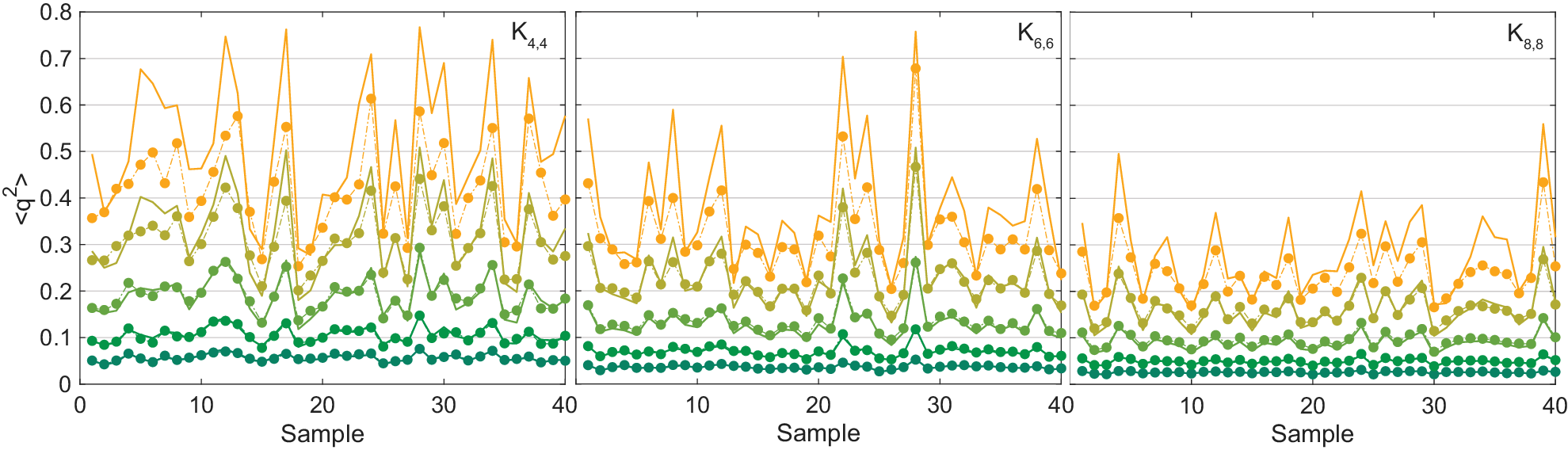}
    \caption{Spin glass order parameter for (from left to right) biclique graphs $K_{4,4}$, $K_{6,6}$ and $K_{8,8}$. The solid lines show the exact result obtained using full state vector simulations with the same Trotterization, while the dots denote the truncated Wigner results. Different colors indicate different annealing times $t_a$ going from $t_a=4$ns (dark green) to $t_a=25$ns (yellow).}
    \label{fig:fluctuationsbenchmark}
\end{figure*}
where $\sigma^{x,z}_i$ denote Pauli matrices acting on qubit $i$, $g(t)$ is the strength of the transverse field and $J(t)$ is the overall strength of the Ising coupling. The Ising interactions $J_{ij}$ are chosen uniform and random on a biclique graphs. That is, for a completely bipartite graph $K_{N,N}$ (composed of $2N$ qubits) we draw uniform random coupling $J_{ij}\sim \mathcal{U}(-\sqrt{4/N},-\sqrt{4/N})$ for every edge connecting the two subsets, as depicted in Fig.~\ref{fig:diagram}.
The goal is to extract the spin glass order parameter 
\begin{equation}
    \left< q^2 \right>=\frac{2}{N(N-1)} \sum_{i \neq j} \left<\sigma^z_i \sigma^z_j \right>^2,
    \label{eq:Q2}
\end{equation}
and the associated Binder cumulant
\begin{equation}
    U=\frac{1}{2} \left(1-\frac{\left< q^4 \right>}{3\left< q^2 \right>^2} \right).
    \label{eq:BinderU}
\end{equation}
One thus needs a method that can accurately extract at least two- and four-point correlation functions.

\emph{Truncated Wigner--}
I will not review the entire literature on phase-space methods, but before presenting the results it is useful to recap the basic ideas behind the truncated Wigner approximation and benchmark it on the problem at hand. Many details can be found in Refs.~\cite{HILLERY1984121,wooters87,polkovnikov10,Schachenmayer15} and references therein. Here we adopt the discrete truncated Wigner approximation as it shows slightly smaller error than the continuous one on our small system benchmarks, but we believe the difference with the continuous Wigner function will become negligible in the large-$N$ limit. In short, truncated Wigner is a semiclassical method based on approximating the time-evolution operator in the Weyl-representation. In practice it amounts to (i) replacing the quantum dynamics generated by Hamiltonian $H$ by classical evolution, and (ii) sampling the initial phase space points according the Wigner distribution of the initial quantum state. For spins the equivalent classical dynamics becomes 
\begin{equation}
    \dot{s}_i^\alpha(t)=\{ s^\alpha_i, H_W\}=2\epsilon_{\alpha\beta\gamma} \frac{\partial H_W}{\partial s^\beta_i} s^\gamma_i(t),
    \label{eq:Eqmot}
\end{equation}
where $\{\cdot,\cdot\}$ denotes the Poisson bracket, $\epsilon_{\alpha\beta\gamma}$ is the Levi-Civita tensor and $H_W$ is the Weyl-ordered Hamiltonian, which is simply
\begin{equation}
 H_W(t)=J(t/t_a) \sum_{i<j}J_{ij} s^z_is^z_j- g(t)\sum_i s^x_i
\end{equation}
In what follows we will always Trotterize the evolution (in timesteps of $dt=0.02$ns), such that the TWA evolution results in a set of consecutive rotations of all the spins around the $x$-axis with angle $\theta^x=-2g(t)dt$ and follwed by a $z$-axis rotation with angle $\theta^z_i=2J(t)\sum_j J_{ij}s^z_j dt$. The computational complexity of the method is thus $O(N^2 t_a)$ and is dominated by the computational of the local field $\theta^z_i$. The initial phase space points $s^\alpha(0)$ are sampled out of the Wigner function, and since we start from an initial $x-$polarized pure state we get
\begin{equation}
    s_i(t=0)=(1, \pm 1,\pm1),
    \label{eq:InitialS}
\end{equation}
i.e. $s_x=1$ and $s_y$ and $s_z$ are chosen $\pm 1$ at random. Such a distribution, not only correctly captures the initial polarization $\left<\sigma^x\right>=1$ but also ensures quantum fluctuations are recovered, e.g. $\left<(\sigma^\mu)^2\right>=1$. Note that this is different from the classical $x$-state $s_i(t=0)=(1,0,0)$. All samples have equal probabilities and positive weights, so one can simply Monte Carlo sample them. This specifies the entire method, to recap, (i) Sample initial phase space points according to Eq.~\eqref{eq:InitialS}, (ii) evolve them in time according to Eq.~\eqref{eq:Eqmot}, and (iii) compute expectation values by averaging the corresponding phase space functions over all the samples. The sampling complexity is similar to that of the actual quantum device. It should be noted that the method has been successfully used to extract correlations and Binder cumulants in long-range Ising models, such as presented in Refs.~\cite{Schachenmayerlongrang15,reyhaneh20}, while high-energy spin glass dynamics has been investigated in Ref.~\cite{silvia20}.

\begin{figure}[t]
    \centering
    \includegraphics[width=0.98\linewidth]{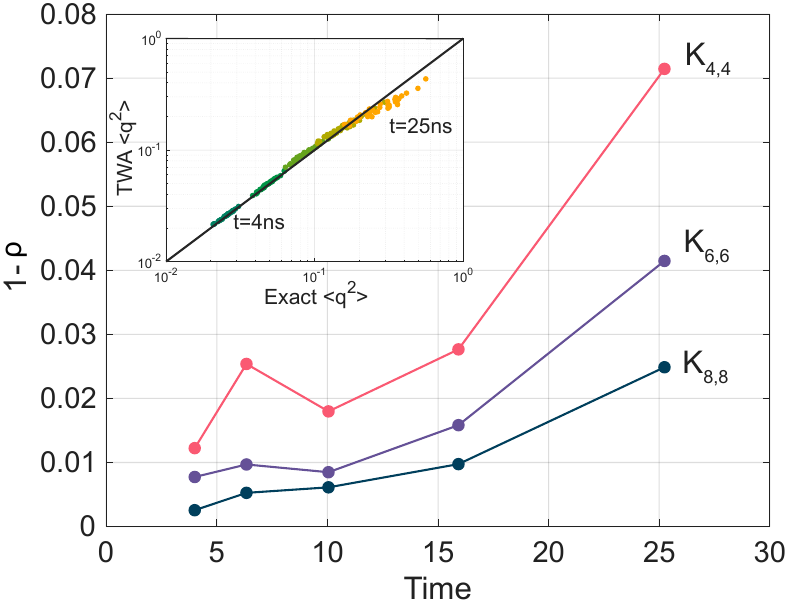}
    \caption{The Pearson correlation coefficient $\rho$ between the TWA and exact $\left<q^2\right>$ is extracted from a population of $N_s=100$ different biclique graphs. As measure of infidelity, the main figure shows $1-\rho$, as a function of annealing time $t_a$, for different system sizes $K_{4,4}$, $K_{6,6}$ and $K_{8,8}$. This inset shows a scatter plot for $K_{8,8}$, and is simply a different representation of the data in Fig.~\ref{fig:fluctuationsbenchmark}; it can be directly compared to Ref.~\cite{dwave25}.}
    \label{fig:Pearson}
\end{figure}

\emph{Benchmark--} Although there are plenty of theoretical reasons to believe the method works well in the large-$N$ limit (see e.g.~\cite{silvia20}), it would be good to asses the accuracy on small systems. In Fig.~\ref{fig:fluctuationsbenchmark} we compare the TWA results to exact state vector simulations on $N_s=40$ different biclique graphs for annealing times from $t=4$ns to $t=25$ns. I would argue that the agreement is excellent, with some expected discrepancies appearing at late times and for samples with large order parameter. To quantify these results we compute the Pearson correlation coefficient $\rho$ over $N_s=100$ disorder realization, as shown in Fig.~\ref{fig:Pearson}. At short, to intermediate, times the correlation is over $99\%$, but then drops at late times resulting in a few percent mismatch on these small systems. Most importantly, the correlation coefficient increases rapidly with system size over the entire range of annealing times, showing TWA captures the relevant finite size fluctuations in the large$-N$ limit.    

Having established the error and scaling on small systems, we extract the Binder cumulant $U$ for biclique graphs ranging from $K_{8,8}$ to $K_{32,32}$ for times ranging from $t_a=4$ns to $t_a=40$ns, thereby simulating the longest times used in Ref.~\cite{dwave25} while simulating larger graphs~\footnote{I don't simulate the dimerization as it seems to be a feature particular to the embedding into D-wave's device, not a feature of the problem. It would be straightforward to add.}. The Binder cumulant is collapsed using the methodology presented in Supplement~IX of Ref.~\cite{dwave25}, and presented in Fig.~\ref{eq:BinderU}. We find a Kibble-Zurek exponent $\mu\approx 5.96$, in line with the theoretical prediction of $\mu=6$. At short times, and for large systems, the cost of estimating the Binder cumulant is entirely dominated by the sampling, as one has to estimate the ratio of the fourth and second moment to very high precision. Note that the same holds true for the quantum experiment. We also extract the Edwards-Anderson order parameters $\left<q^2\right>$, and collapse it using the previously extract exponent $\mu$, finding an anomalous exponent of $r=0.036$. Since the order parameter can be estimated with fewer samples we also add $K_{64,64}$ and $K_{128,128}$ to Fig.~\ref{eq:BinderU} to highlight the computational efficiency of the method. 

\emph{Discussion--} In line with seminal results from the 90's on the equilibrium properties of the quantum Ising spin glass transition~\cite{ye93,miller93}, we numerically establish that the truncated Wigner approximation captures the non-equilibrium properties of the transition. We stress that TWA captures the leading finite size effects, revealed in the sample-to-sample fluctuations of the order parameter. If one desires greater accuracy, one can straightforwardly extend these methods using cluster truncated Wigner approximation as introduced in Ref.~\cite{wurtz18}. In addition, Ref.~\cite{wurtz18} presents a scheme to ``purify'' the cluster operators, requiring only sampling and evolving cluster wave functions. This suggests a scheme to combine tensor network methods with phase space sampling for heterogeneous graphs.    
\begin{figure}[t]
    \centering
    \includegraphics[width=0.88\linewidth]{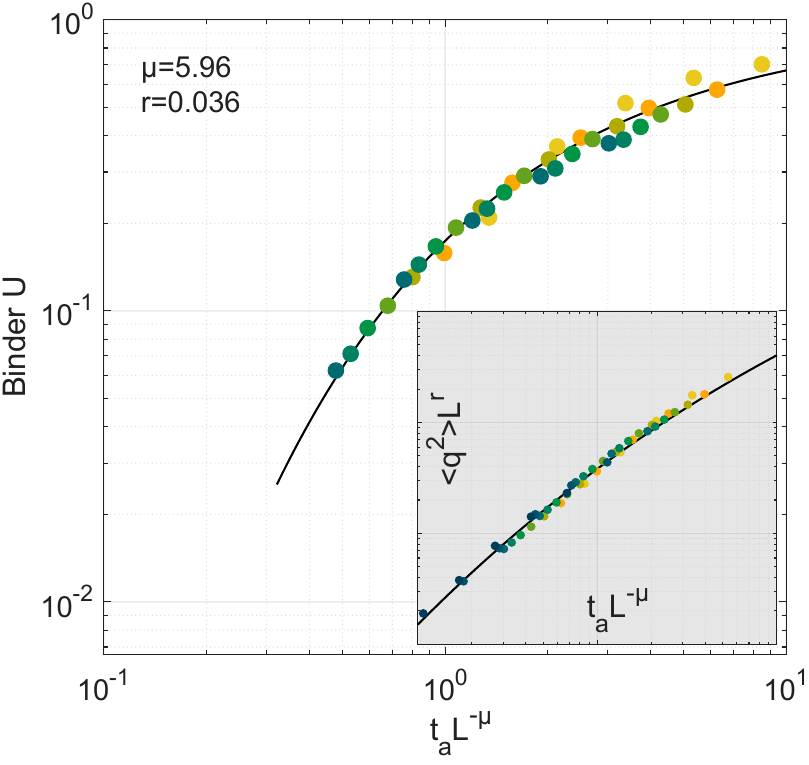}
    \caption{The main figure shows the Binder cumulant $U$ extracted as a function of rescaled time, for systems ranging from $K_{8,8}$ (yellow) to $K_{32,32}$ (blue) (increasing each subset by 4 qubits) and times ranging from $t_a=7$ns to $t_a=40$ns. The inset shows the rescaled order parameter, adding additional $K_{64,64}$ and $K_{128,128}$ graphs.}
    \label{fig:BinderUQ}
\end{figure}

These results do not simply challenge quantum supremacy claims regarding biclique graphs, in conjunction with other results, they seriously constrain the prospect of practical quantum advantage in solving optimization problems. Reference~\cite{watanabe2026tensornetworksurrogatemodels} shows that quantum approximate optimization algorithms (QAOA) tends to drive the system to low entangled states when optimized at sufficiently large depths. In Ref.~\cite{morone2026variationaliterativerotationalgorithm} it was shown that a classical version of QAOA outperform the quantum algorithm on the SK spin-glass problem at any depth. Essentially leaving us only with some very non-local, yet non-mean-field problems, for which error correction is completely unrealistic~\cite{stoudenmire24}.

\emph{Acknowledgements}
The Flatiron Institute is a division of the Simons Foundation. I am supported by AFOSR under Award No. FA9550-21-1-0236. I thank A. Polkovnikov and J. Tindall for useful discussions and the facilities
of the Boston University SCC. 

\clearpage
\bibliography{library}

\end{document}